\documentstyle [12pt] {article}

\begin{document}
\begin{flushright}
IUHET-264\\
October 1993\\
\end{flushright}
\vspace*{0.5 cm}
\begin{center}
{\bf
CONSTRAINTS ON THREE-NEUTRINO MIXING\\
FROM ATMOSPHERIC AND REACTOR DATA
}\\
\vspace*{0.5 cm}
\vspace*{0.5 cm}
{\bf  J. Pantaleone}\\
\vspace*{0.5 cm}
Physics Department\\
Indiana University, Bloomington, IN 47405 \\
\vspace*{0.5 cm}
{\bf ABSTRACT} \\
\end{center}

Observations of atmospheric neutrinos are usually analyzed using the
simplifying approximation that either \(\nu_\mu \leftrightarrow \nu_\tau\)
or \(\nu_e \leftrightarrow \nu_\mu\) two-flavor mixing is relevant.
Here we instead consider the data using the simplifying
approximation that only one neutrino mass scale is relevant.
This approximation is the minimal three-flavor notation
that includes the two relevant two-flavor approximations.
The constraints in the parameter space orthogonal to the
usual, two-flavor analyses are studied.

\newpage

In recent years, large water Cherenkov detectors located
deep underground have been able to provide statistically significant
measurements of the flux of atmospheric neutrinos \cite{K,IMB}.
They have found that the flavor content of the flux differs
from expectations \cite{flux}.
In particular, Kamiokande and IMB have found the ratio of
\(\nu_\mu / \nu_e \) in their fully contained events to be
0.60 \(\pm\) 0.09 and 0.55 \(\pm\) 0.09 , respectively, of what they expected.
Smaller, but differently constituted, detectors have provided less
statistically significant results that are consistent with these
observations \cite{Frejus,S}.  This discrepancy could be caused by
neutrino mixing (see e.g. \cite{LPW,Beier}).

Atmospheric neutrinos involve measurably distinct
fluxes of more than one flavor of neutrino.  In principle,
they depend on the full possible range of
neutrino mixing parameter space: two mass-squared difference scales,
three mixing angles and a CP violating phase.
In practice, the present data are rather crude
so that the presence of any nonzero neutrino oscillations parameters
is not yet certain.
Thus the observations have (generally) only been analyzed in the two-flavor
approximation where the number of parameters is minimal:
one mass scale and one mixing angle.
However there are good reasons for going beyond the two-flavor
approximation.

One reason for going beyond the two-flavor approximation is that
the data presently indicate that at least one of the mixing
angles is rather large \cite{FGMS}.
Also, the effective \(\nu_e\) mixing is sometimes
enhanced for atmospheric neutrinos by matter effects.
As is well known, the two-flavor
approximation can be quite poor when a mixing angle,
either a vacuum angle or a matter enhanced angle \cite{KP3,KP,P}, is large.

Neutrino masses are small because of the structure of the standard model
\cite{Weinberg}.  Neutrino mixings can be predicted in some
extensions of the standard model.
At present, the most attractive extension of the standard model
is the SO(10) Grand Unified Theory.
In this model, with ``minimal'' particle content,
the neutrino mixing is calculable in terms of the quark
mixing---however both small and large mixing solutions exist \cite{Lavoura}.
Additional, unpredictable contributions are generally also present.
Thus theoretical arguments can not exclude any values of the
vacuum mixing angles.

A second reason for going beyond the two-flavor approximation is
to understand how to test the neutrino oscillation explanation of the
measurements.
The continuing experiments and many new experiments presently
under construction \cite{newatm} will improve our knowledge of
the atmospheric neutrino flux.  However uncertainties in the
calculation of the atmospheric neutrinos production limit
the ability of these experiments to constrain neutrino mixing parameters.
Consequently many new experiments using neutrinos
produced at reactors \cite{newrea} and at accelerators \cite{newacc}
are being planned to study the specific neutrino parameters
believed to be relevant for the atmospheric neutrino discrepancy.
Hence it is important to know exactly what those parameters are.

A third reason for going beyond the two-flavor approximation is
because the contained atmospheric events involve more than one
flavor of neutrino.
Hence they can be explained by two different types of two-flavor mixing,
either \(\nu_\mu \leftrightarrow \nu_\tau\) or
\(\nu_e \leftrightarrow \nu_\mu\).  Analyses of the data are
routinely done in both of these approximations.
However, as is generally realized but seldom discussed,
there is a continuous parameter space between these two limits.
The purpose of this article is to clarify and explore this intermediate
region between the two relevant two-flavor approximations.

Here we compromise and examine the data in a simplified
three-flavor formalism.  We {\it assume} that one of the
mass-squared scales is less than \(3 \times 10^{-5}\)  eV\({^2}\)
and hence irrelevant since the associated
oscillation wavelength is longer
than the longest propagations lengths of the current experiments
(the diameter of the Earth) \cite{SNP}.
Then there is only one mass-squared scale
and two mixing angles which are relevant.
A heuristic way to express this is to just set
\begin{equation}
m_1 = m_2 = 0,  \  \ m_3 > 0 .
\end{equation}
Then the remaining two mixing angles
can be thought of as defining the amount of \(\nu_e\) and \(\nu_\mu\)
in the one massive state
(the amount of \(\nu_\tau\) in this state is fixed by unitarity).

The parameterization of the mixing
\begin{equation}
| \nu_\alpha > \ \ = \ \ U_{\alpha i} | \nu_i >
\end{equation}
between the flavor eigenstates, \(\alpha = e, \mu, \tau\),
and the mass eigenstates, \(i = 1, 2, 3\),
is here chosen to be
\begin{equation}
U = \left[ \begin{array}{ccc}
0 &  \cos \phi & \sin \phi \\
- \cos \psi & - \sin \psi \sin \phi & \sin \psi \cos \phi \\
  \sin \psi & - \cos \psi \sin \phi & \cos \psi \cos \phi
\end{array} \right]
\label{U}
\end{equation}
where \(\phi\) and \(\psi\) are the mixing angles.
This parameterization is chosen
such that matter effects \cite{W} are straightforward
(for a general review of matter effects see \cite{KP}).
In a matter background, \(\psi\) is unchanged but the
effective \(\phi\) is given by
\begin{equation}
\sin^2 2 \phi_m = { { (m_3^2 \sin 2 \phi )^2 } \over
{ (A-m_3^2 \cos 2 \phi )^2 + (m_3^2 \sin 2 \phi )^2 } }  \  \ .
\end{equation}
and the effective mass eigenstates are
\begin{eqnarray}
M_1^2 &=& 0 \\
M_{3,2}^2 &=& {1 \over 2} [ ( m_3^2 + A ) \pm
\sqrt{ (A-m_3^2 \cos 2 \phi )^2 + (m_3^2 \sin 2 \phi )^2 } ]
\nonumber
\label{M}
\end{eqnarray}
where the i=2 state is associated with the minus sign \cite{order}.
Here \(A\) is the induced mass-squared from the electron background,
\begin{eqnarray}
A &=& 2 \sqrt{2} G_F ( Y_e \rho / m_u ) E  \\
  &=& 3.8 \times 10^{-4} eV^2
\left( { Y_e \rho \over 2.5 g/cm^3 } \right)
\left( { E \over 1 GeV } \right)
\nonumber
\label{A}
\end{eqnarray}
with \(G_F\) as Fermi's constant,
\(Y_e\) is the number of electrons per nucleon,
\(\rho\) is the density, \(m_u\) is the nucleon mass,
and \(E\) the neutrino energy.
For antineutrinos, \(A \rightarrow - A\).

To illustrate the physical implications of this parametrization,
we give the relevant oscillations probabilities
for a constant density medium.
\begin{eqnarray}
P(\nu_e \rightarrow \nu_e) &=&
1 - {1 \over 2} \sin^2 2 \phi_m [1 - \cos ( \beta_3 - \beta_2 ) ]
\nonumber \\
P(\nu_\mu \rightarrow \nu_e) &=&
{1 \over 2} \sin^2 \psi \sin^2 2 \phi_m [1 - \cos ( \beta_3 - \beta_2 ) ]
\\
P(\nu_\mu \rightarrow \nu_\mu) &=&
1  - {1 \over 2} \{ \sin^2 2 \psi [ 1
- \sin^2 \phi_m \cos (\beta_2 - \beta_1)
- \cos^2 \phi_m \cos(\beta_3 - \beta_1) ] \nonumber \\
& & + \sin^4 \psi \sin^2 2 \phi_m [ 1 - \cos ( \beta_3 - \beta_2 ) ] \}
\nonumber
\label{P's}
\end{eqnarray}
Here the dynamical phase acquired by a neutrino mass eigenstate which
propagates for a time t is
\begin{equation}
\beta_i \equiv { { M_i^2 t } \over {2 E } }
\end{equation}
Unitarity and time reversal symmetry \cite{KPCP}
can be used to obtain the other oscillation
probabilities from those above.
For more complicated density distributions,
the procedure for calculating the probabilities
is straightforward (see e.g. \cite{KP}).

In the two-flavor {\it vacuum} approximation,
neutrino oscillation effects are symmetric for mixing angles
in the ranges 0 to \(\pi/4\) and \(\pi/4\) to \(\pi/2\).
However when there are three flavors (and also when matter effects
are relevant) there is no symmetry between these two ranges.
Thus we use limits where  \(\phi \) and \(\psi\) explicitly range
between 0 and \(\pi/2\), or, equivalently, the sines of these angles
range between 0 and 1.
This covers the full allowed range for these parameters,
without any redundancy.

When one of the mixing angles is at the limit of its range,
then one of the neutrino flavors decouples and
the approximate three-flavor notation
(Eq. (\ref{U})) reduces to a two-flavor
description in terms of the remaining mixing angle (see Table).
All possible two-flavor approximations are included.
However since there are  only two mixing angles,
the three possible types
of two-flavor approximation are not fully independent.
This just follows from the assumption of
only one relevant mass scale.

In this notation the neutrino parameter space has three dimensions:
two mixing angles and the mass-squared.  The conventional
two-flavor plots of mass-squared versus a mixing angle correspond to
one of the two dimensional surfaces of this three dimensional parameter space.
To complement this usual approach, we here show plots (Figs. (1) and (2))
at fixed mass-squared.
These plots of one mixing angle versus the other mixing angle
show cross sections of
the parameter space which are orthogonal to the usual two-flavor plots.

The region between the dotted contours in Fig. (2) is
excluded by reactor measurements \cite{IRP}.
Reactor experiments are purely \(\nu_e\) disappearance experiments.
As can be seen from Eqs. (7),
the expression for \(P ( \nu_e \rightarrow \nu_e )\) is
always equivalent to the two-flavor approximation.
Thus no additional calculations are necessary.
The mass-squared value in Fig. (1) is below
the present two-flavor limits from reactor experiments,
so there are no constraints.
For the fixed mass-squared of Fig. (2),
the mixing angle is a fixed value
and the dotted contour is a vertical line.

Atmospheric neutrinos involve more than one flavor,
so the three-flavor effects are significant
and the constraint contours must be calculated accordingly.
Following the accepted two-flavor analyses,
we calculate oscillation constraints using
ratios of atmospheric neutrino flux measurements.
This is to cancel out the large errors inherent in modeling the production
of atmospheric neutrinos.  Two different ratios are used:
\(R_\nu\), the ratio of \(\nu_\mu\) and
\(\nu_e\) fluxes in fully contained events (\(<E> \sim\) 0.8 GeV);
and \(R_\mu\), the ratio of the fluxes of upward going muons that
stop in the detector (\(<E> \sim\) 10 GeV)
to those that go completely through the detector
(\(<E> \sim\) 100 GeV).  For both of these ratios,
we do not perform a chi-squared fit to
the energy and angular distributions of the data
(which would require detailed knowledge of the detector resolutions,
efficiencies, thresholds, scattering cross sections, etc),
but instead just fit to the total flux ratios.
For \(R_\nu\) we use Kamiokande's value, for \(R_\mu\) we use
IMB's value with an extra 7\% systematic uncertainty \cite{FGMS}.
The ratio of \(\nu_\mu / \nu_e\) at production is taken to be 2.0,
the ratio of \({\bar \nu} / \nu\) in the detected events is taken to be 0.4,
the oscillation probabilities are averaged over the approximate
energy distributions given in ref. \cite{FGMS} and a 1/L neutrino
path length distribution.   The contours shown in the Figures are all
90\% confidence levels.

In the Figures,
the parameter regions allowed by all the constraints are shaded.
The parameter region preferred by
\(R_\nu\) ranges continuously from the left boundary
(\(\nu_\mu \leftrightarrow \nu_\tau\) mixing)
to the upper boundary (\(\nu_e \leftrightarrow \nu_\mu\) mixing).
The parameter region excluded by \(R_\mu\)
connects only to the left boundary (\(\nu_\mu \leftrightarrow
\nu_\tau\) mixing).
The values on the boundaries are in rough agreement with previous
two-flavor analyses.
The constraint by \(R_\mu\) in the \(\nu_e \leftrightarrow \nu_\mu\) two-flavor
approximation has not been discussed previously by the experimental groups.
That limiting constraint is sensitive to the ratio of \(\nu_\mu / \nu_e\) at
high energies, and it vanishes for the ``large'' value of
this ratio used here.

Somewhat below the mass-squared value of Fig. (1), the average
oscillation wavelength becomes larger than the radius of the Earth
and all of the constraints vanish.
The constraints from \(R_\mu\) are smaller at the larger mass-squared
of Fig. (2), and quickly vanish with increasing mass-squared,
because then both the stopping and through-going muon fluxes are equally
reduced by mixing.  The constraints from \(R_\nu\) are
constant with increasing mass-squared.

The constraints in the Figures were also computed
without the matter background, to study the importance of this effect.
For the \(R_\nu\) contours, the value of \(\sin^2 \phi\) near the center
of the Figures would be reduced by 50\% if matter effects were neglected,
while the values on the boundaries would essentially remain unchanged.
For the \(R_\mu\) contours,
the excluded area would be reduced by a third to a half if
matter effects were neglected.
Thus the matter background is quantitatively quite
important for the atmospheric neutrino contours.

The Figures illustrate some general features of three-flavor effects.
For example, note that the contours
from the reactor experiments and \(R_\mu\)
are mostly vertical and horizontal (on the left) lines,
respectively, while the \(R_\nu\) contours are sloping.
Thus the overlap between the two excluded regions and the preferred
region is rather imperfect.
An enlargement of these exclusion regions could result
in a situation where, at a given mass-squared,
the two two-flavor approximations showed no allowed region
while there still was some allowed region in a three-flavor analyses
(toward the upper left corner of the Figures).
Thus we can conclude that to definitively exclude
the possibility of an allowed region, the constraint contours
may have to extend well beyond what
was indicated by the two-flavor analyses.

Appearance experiments are typically far more sensitive
to neutrino oscillations than disappearance experiments.
A long-baseline \(\nu_\mu\) to \(\nu_e\) appearance experiment
is experimentally feasible, and would have a large overlap
with the three-flavor allowed region from contained atmospheric events.
In addition, matter effects enhance the sensitivity
of such experiments, as noted previously \cite{P}.

In summary, the one mass scale approximation gives a simple,
reasonable, three-flavor notation for analyzing atmospheric
and reactor neutrino observations.
There are then only three neutrino parameters:
one mass scale and two mixing angles.
The parametrization here smoothly describes the transition between the
the \(\nu_\mu \leftrightarrow \nu_\tau\) and \(\nu_e \leftrightarrow \nu_\mu\)
two-flavor approximations usually used to analyze the data.
To complement the usual calculations,
the constraints at fixed mass squared have been computed
from atmospheric neutrino contained events,
atmospheric neutrino induced muons, and reactor neutrinos.
The matter background of the Earth has been included,
and has a quantitatively important effect on the contours.
A three-flavor notation may be crucial for determining
atmospheric neutrino oscillations.

\begin{center}
Acknowledgements
\end{center}

I would like to thank R. Heinz, A. Hendry and S. Mufson
for useful discussions.  This work is supported in part by
the U.S. Department of Energy under Grant No. DE-FG02-91ER40661.

\raggedbottom

\raggedbottom
\newpage

Table.
The two mixing angles, \(\psi\) and \(\phi\),
range between 0 and \(\pi/2\).  When one of these mixing
angle is at the limit of its range,
this three-flavor notation (Eq. (\ref{U})) reduces to
a two-flavor approximation.
The parameter limits and corresponding equivalent
two-flavor approximation are given below.
\\

\begin{tabular}{| c | c |} \hline
Angle limit  & Equivalent two-flavor mixing \\ \hline
\(\sin^2 \psi = 1.0\)  & \(\nu_e \leftrightarrow \nu_\mu\)   \\
\(\sin^2 \psi = 0.0\)  & \(\nu_e \leftrightarrow \nu_\tau\)   \\
\(\sin^2 \phi = 0.0\) & \(\nu_\mu \leftrightarrow \nu_\tau\)  \\
\(\sin^2 \phi = 1.0\) & no oscillations \\
\hline
\end{tabular}

\raggedbottom
\newpage

\vspace*{0.6cm}

\noindent {\bf Figures.}
Plots of \(\sin^2 \psi\) versus \(\sin^2 \phi\) at constant
mass-squared.
The dashed lines show the excluded
region from the ratio of stopping/through-going
atmospheric neutrino induced muons.
The dotted lines show the excluded region from reactor experiments.
The solid line shows the preferred region from contained
atmospheric neutrino observations.
The allowed region is shaded.
(1) \(m^2 = 3 \times 10^{-3}\) eV\(^2\),
(2) \(m^2 = 3 \times 10^{-2}\) eV\(^2\).

\raggedbottom


\begin{thebibliography}{99}
\bibitem{order}
Taking the \(m_2 \rightarrow m_1\) limit of the general
notation of refs. \cite{KP3,KP} yields this notation
when the 1 and 2 indices are interchanged.
\end{thebibliography}
\end{document}